# EPR spectra and magnetization of XY-type rare-earth ions in pyrochlores $Y_2Ti_2O_7$:$R^{3+}$ (R=Yb, Er)

R.G. Batulin[1], M.A. Cherosov[1], I.F. Gilmutdinov[1], A.G. Kiiamov[1], V.V. Klekovkina[1,*], B.Z. Malkin[1], A.A. Rodionov[1], R.V. Yusupov[1]

[1] Kazan Federal University, Kremlevskaya 18, Kazan 420008, Russia
*E-mail: Vera.Klekovkina@kpfu.ru

The results of studies of $Y_2Ti_2O_7$ single crystals doped with $Er^{3+}$ or $Yb^{3+}$ ions by means of electron paramagnetic resonance (EPR) and dc-magnetometry are reported. EPR signals of the trigonal centers with the characteristic hyperfine structure of $Er^{3+}$ or $Yb^{3+}$ ions were observed. Field dependences of the magnetization of single crystals for magnetic fields directed along the crystallographic axes and temperature dependences of the magnetic susceptibilities were measured. Spin Hamiltonian parameters (g-factors and parameters of the hyperfine interaction) for the $Er^{3+}$ and $Yb^{3+}$ ions were obtained from the analysis of the experimental data. The registered EPR spectra and magnetization curves are successfully reproduced by simulations in the framework of the crystal-field approach, in particular, with an account for a hybridization of the ground $4f^{13}$ configuration of $Yb^{3+}$ ions with the charge transfer states.

## 1. Introduction

Extensive experimental and theoretical studies of the magnetic structures and excitations in the so called XY pyrochlores containing rare-earth (RE)$Er^{3+}$ or $Yb^{3+}$ ions with the easy-plane magnetic anisotropy have been carried out over the last twenty years ([1-3] and references therein). A diversity of magnetic phases in the concentrated materials $Yb_2M_2O_7$ and $Er_2M_2O_7$ (M=$Ge^{4+}$, $Ti^{4+}$, $Pt^{4+}$, $Sn^{4+}$) was revealed at low temperatures (noncoplanar $\psi_2$ or coplanar $\psi_3$ antiferromagnetic order [4-7], antiferromagnetic phase of Palmer-Chalker [9,10], splayed ferromagnetic order [10-14]).

RE ions in the crystal lattice of the pyrochlore (space group $Fd\bar{3}m$) occupy 16d Wyckoff positions with the local trigonal symmetry $D_{3d}$. Four magnetically non-equivalent RE sublattices form a corner-sharing tetrahedral network, the basis vectors of $RE^{3+}$ ions in the unit cell in the crystallographic system of coordinates with the origin at the center of a corresponding tetrahedron are as follows: $\mathbf{r}_1 = (1,1,1)/8$, $\mathbf{r}_2 = (-1,-1,1)/8$, $\mathbf{r}_3 = (-1,1,-1)/8$, $\mathbf{r}_4 = (1,-1,-1)/8$ in the units of the lattice constant $a$. The first coordination shell of RE ions contains eight oxygen ions which form strongly distorted cubic polyhedron with the two nearest neighbor oxygen ions at one of the four crystallographic [111] axes. Specific magnetic properties of different RE pyrochlores are determined, first of all, by the single RE ion magnetic characteristics and the energy spectrum in the trigonal crystal field (CF).

The wave functions of the ground Kramers doublet of $Er^{3+}$ or $Yb^{3+}$ ions in the trigonal CF transform accordingly to the irreducible representations $\Gamma_4$ or $\Gamma_{56}$ of the $D_{3d}$ point symmetry group. Note that magnetic properties of these states are substantially different, in particular, the so called dipole-octupole doublets $\Gamma_{56}$ are split only by the magnetic field parallel to the trigonal symmetry axis (the transversal g-factor $g_\perp = 0$) while both the longitudinal ($g_\parallel$) and the transversal g-factors of the $\Gamma_4$ doublets are non-zero.

The CF splittings of the electronic $4f^N$ – multiplets and spectroscopic g-factors of $Er^{3+}$ (N=11) and $Yb^{3+}$ (N=13) ions in dilute and concentrated pyrochlores were extensively studied earlier. The g-factors of the ground state of $Yb^{3+}$ ions were determined from $^{170}Yb$ Mössbauer absorption measurements in $Y_2Ti_2O_7$:$Yb^{3+}$ (1 at.%) sample enriched with the $^{170}Yb$ isotope in Ref. 15, but g-factors of $Er^{3+}$ ions were estimated only from calculations based on the CF parameters which had been used for simulations of the dc-magnetic susceptibility of$Er_2Ti_2O_7$ [16]. Some CF energies of $Yb^{3+}$ ions in $Y_2Ti_2O_7$:$Yb^{3+}$ and $Yb_2Ti_2O_7$ were measured by means of optical spectroscopy [17]. CF excitations corresponding to transitions between sublevels of the ground multiplets of $Yb^{3+}$ and $Er^{3+}$ in $Yb_2Ti_2O_7$ and $Er_2M_2O_7$ (M=Ge, Ti, Sn, Pt, Ru) were studied by means of inelastic neutron scattering in Refs. [18,19] and [20,21], respectively.

The CF approach serves as a basis for construction of theoretical models of interactions between RE ions and a crystal lattice and for understanding of magnetic properties of concentrated RE compounds. A number of CF parameter sets for $Er^{3+}$ and $Yb^{3+}$ in the pyrochlores with different chemical compositions were proposed in the literature (see [1,22,23] and references therein). However, the comment written 18 years ago [15], namely, "All are inappropriate as each corresponds to a crystal level scheme and to the ground state wave functions which are not compatible with the experimental data", is actual nowadays as well. Determination of the physically consistent CF parameters describing both the measured g-factors and CF energies remains a topical problem.

In recent work [24], stimulated by the increasing interest to the unconventional magnetic properties of geometrically frustrated RE pyrochlores, we presented the measured electron paramagnetic resonance (EPR) and site-selective emission and excitation optical spectra of $Y_2Ti_2O_7$ single crystals doped with $Er^{3+}$ or $Yb^{3+}$ ions (0.5 at.%). The CF parameters were calculated in the framework of the semi-phenomenological exchange charge model and then corrected to fit the experimental data. The main goal of the present work is to obtain additional information about electronic structures of pyrochlores containing XY-type RE ions ($Er^{3+}$ or $Yb^{3+}$). We present low-temperature magnetic field dependences of the magnetization and temperature dependences of the bulk dc-susceptibility of $Y_2Ti_2O_7$:$Yb^{3+}$ and $Y_2Ti_2O_7$:$Er^{3+}$ (0.5 at. %) single crystals. The shape and the hyperfine structure of EPR signals presented in [24] and magnetization curves are analyzed by making use of the corresponding sets of CF parameters.

## 2. Experimental details and results

The measurements of the magnetization $M(\mathbf{B},T)$ parallel to the external magnetic field $\mathbf{B}$ with magnetic fields in the range of 0−9 T applied along the crystallographic axes [100], [111] and [110] were performed using a vibrating sample magnetometer option of the PPMS-9 system (Quantum Design) on $Y_2Ti_2O_7$:$Yb^{3+}$ and $Y_2Ti_2O_7$:$Er^{3+}$ single crystals with the content of 0.5 at.% of impurity ions grown by the optical floating-zone method [25]. The details of the single crystal growths were described in [24]. The pyrochlore structure of the samples was confirmed by X-ray diffraction measurements. The obtained magnetic field dependences of the magnetization at the temperature $T$ of 2 K are shown in Figure 1. The experimental data in Figure 1 are corrected by accounting for diamagnetic contributions [26] into the measured magnetization which are substantial in case of strongly dilute samples.

For weak magnetic fields, $B<1$ T, the magnetization $\mathbf{M}(\mathbf{B},T)$ is practically isotropic, as one may expect in the case of global cubic symmetry of the studied systems. The temperature dependences of the static magnetic susceptibility $\chi(T)=M(B,T)/B$ of $Y_2Ti_2O_7$:$Yb^{3+}$ and $Y_2Ti_2O_7$:$Er^{3+}$ single crystals measured in the magnetic fields $B$=0.4 T and 0.3 T, respectively, are presented in Figure 2. The measured susceptibility curves can be satisfactorily approximated by the Curie law, $\chi(T)=C/T$, with the Curie constant $C=2cN_A g^2 J(J+1)\mu_B^2/3k_B$ at elevated temperatures and $C=cN_A(g_\parallel^2+2g_\perp^2)\mu_B^2/6k_B$ at low temperatures (here $c$ is the concentration of RE ions, $N_A$ is the Avogadro number, $\mu_B$ is the Bohr magneton, $k_B$ is the Boltzmann constant, $g$ is the Lande factor of the ground multiplet with the total angular moment $J$, $g_\parallel$ and $g_\perp$ are the g-factors of the ground Kramers doublet determined from the EPR spectra (see below)). As the magnetic field increases, a noticeable dependence of the magnetization on the field direction appears due to mixing of wave functions of the ground and excited CF levels by the Zeeman interaction (see Figure 1). Of particular interest for justification of the CF parameter sets is the difference in the relative shifts of magnetization curves of $Er^{3+}$ and $Yb^{3+}$ ions in magnetic fields along the tetragonal [001] axis, in this case the observed magnetization in strong magnetic fields has maximum values for the $Yb^{3+}$ ions but minimum values for the $Er^{3+}$ ions. The gaps of ~600 $cm^{-1}$ and ~50 $cm^{-1}$ between the ground $\Gamma_4$ doublet and the first excited state of $Yb^{3+}$ and $Er^{3+}$ ions[24], respectively, differ by more than an order of magnitude, and, in agreement with the nonlinear mechanism of the single-ion magnetic anisotropy in cubic paramagnetic centers, it is remarkably stronger in $Y_2Ti_2O_7$:$Er^{3+}$ than in $Y_2Ti_2O_7$:$Yb^{3+}$ single crystals.

The EPR spectra were measured with the commercial Bruker ESP300 X-band spectrometer in static magnetic fields up to 550 mT. The registered spectra of the $Y_2Ti_2O_7$:$Yb^{3+}$ and $Y_2Ti_2O_7$:$Er^{3+}$ (0.5 at.%) single crystals with the magnetic fields directed along the crystallographic axes are shown in Figures 3 and 4, respectively. Using the measured resonant magnetic fields of even isotopes, we found the corresponding effective $g$-factors. The spectra measured with the magnetic field **B** directed along the tetragonal [001] axis contain one intense line corresponding to the $g$-factor $g_{[001]} = \left(g_\parallel^2/3 + 2g_\perp^2/3\right)^{1/2}$ of the four magnetically equivalent paramagnetic centers (here and below $g_\parallel$ and $g_\perp$ are $g$-factors in the local coordinate frames with the $Z$-axis along the trigonal symmetry axis of the corresponding center). If the external magnetic field is parallel to a trigonal [111] axis, there are two magnetically non-equivalent centers of $RE^{3+}$ ions: one site whose local anisotropy axis is parallel to the static magnetic field **B** and three sites have their local anisotropy axis at ~109.5 degrees from the vector **B**. In the magnetic field applied along the two-fold rhombic [110] axis, there are also two magnetically non-equivalent centers: two so-called $\alpha$-sites (the magnetic field is declined from the local trigonal axis by the angle of ~35.3 degrees) and two $\beta$-sites (the field is perpendicular to the local trigonal axis). Therefore in cases of static magnetic field directed along trigonal or rhombic axes, two EPR lines with different intensities corresponding to the $g$-factors $g_{[111]}^{(1)} = g_\parallel$, $g_{[111]}^{(2)} = \left(g_\parallel^2/9 + 8g_\perp^2/9\right)^{1/2}$, and $g_{[110]}^{(1)} = \left(2g_\parallel^2/3 + g_\perp^2/3\right)^{1/2}$, $g_{[110]}^{(2)} = g_\perp$ are observed. The sum of the squared g-factors over four ion sites in the unit cell for each direction of the magnetic field that determines the linear magnetic susceptibility is a constant. Along with the most intensive signals from even isotopes, the spectra contain weaker components of the hyperfine structure originating from odd isotopes ($^{167}$Er with the nuclear spin $I$=7/2 and natural abundance of 22.9 %, $^{171}$Yb (14.3 %) and $^{173}$Yb (16.2 %) with $I$=1/2 and $I$=5/2, respectively). The measured values of the $g$-factors, $g_\parallel$ and $g_\perp$, and the hyperfine structure parameters for odd isotopes are presented in Table 1. We note that the $g$-factors of $Er^{3+}$ ions in the concentrated Er-pyrochlore $Er_2Ti_2O_7 g_\parallel$=2.6 from Ref. 27 (polarized neutron diffraction measurements) and $g_\perp$=7.6 from Ref. 28 (low-temperature EPR transmission-type spectroscopy) are rather far from our values.

3. Discussion

In order to simulate the EPR spectra, magnetization and susceptibility data, we considered the following Hamiltonian of a rare-earth ion,

$$H = H_{FI} + H_{CF} + H_Z + H_{HF}. \qquad (1)$$

Here $H_{FI}$ is the free ion standard parameterized Hamiltonian [29] that operates in the total space of states of the electronic $4f^N$ configuration (364 and 14 states for the $Er^{3+}$ and $Yb^{3+}$ ions, respectively). $H_{CF}$ is the CF energy of $4f$ electrons, $H_Z$ is the electronic Zeeman energy, and $H_{HF}$ corresponds to the hyperfine interaction.

The local systems of coordinates defined in [30] for RE ions at sites with the radius-vectors $\mathbf{r}_n$ are used in calculations: the $Z_n$ axes are parallel to $\mathbf{r}_n$, and the $X_n$ axes are in the planes containing $\mathbf{r}_n$ and the selected crystallographic tetragonal axis coinciding with the $z$-axis of the global coordinate frame. The CF Hamiltonian written in the local system of coordinates,

$$H_{CF} = \sum (B_2^0 O_2^0 + B_4^0 O_4^0 + B_4^3 O_4^3 + B_6^0 O_6^0 + B_6^3 O_6^3 + B_6^6 O_6^6), \qquad (2)$$

is determined by six non-zero CF parameters $B_p^k$ (here $O_p^k$ are linear combinations of the single-electron spherical tensor operators [22]), the sum is taken over $4f$ electrons with radius-vectors **r**, orbital and spin moments **l** and **s**, respectively. The values of the CF parameters used to simulate the EPR spectra, the dc-susceptibility and the magnetization data are given in Table 2. The operator corresponding to the Zeeman energy has the form $H_Z = -\mathbf{\mu B}$, where $\mathbf{\mu} = -\mu_B(k\mathbf{L}+2\mathbf{S})$ is the magnetic moment of a RE ion, **L** and **S** are orbital and spin moments, respectively, and $k$ is the orbital reduction factor ($k$=0.99 for the $Er^{3+}$ ion and 0.98 for the $Yb^{3+}$ ion).

The Hamiltonian $H_{HF}$ of hyperfine interactions contains magnetic ($H_{HFM}$) and quadrupole ($H_{HFQ}$) contributions:

$$H_{HFM} = \mu_B \gamma_N \hbar \langle r^{-3} \rangle_{4f} \sum \{2\mathbf{lI} + O_2^0(3s_zI_z - \mathbf{sI}) + 3O_2^2(s_xI_x - s_yI_y) + 3O_2^{-2}(s_xI_y + s_yI_x) \\ + 6O_2^1(s_zI_x + s_xI_z) + 6O_2^{-1}(s_yI_z + s_zI_y)\}, \quad (3)$$

$$H_{HFQ} = \frac{e^2Q}{4I(2I-1)}\{(1-\gamma_\infty)\sum_L q_L \frac{3z_L^2 - r_L^2}{r_L^5}(3I_z^2 - I(I+1)) - (1-R_Q)\langle r^{-3} \rangle_{4f} \sum \left[O_2^0(3I_z^2 - I(I+1))\right. \\ \left. + 3O_2^2(I_x^2 - I_y^2) + 3O_2^{-2}(I_xI_y + I_yI_x) + 6O_2^1(I_zI_x + I_xI_z) + 6O_2^{-1}(I_yI_z + I_zI_y)\right]\}. \quad (4)$$

Here $\gamma_N$ is the nuclear gyromagnetic ratio ($\gamma_N/2\pi$ =−1.18, 6.9 and −1.98 MHz/T for $^{167}Er$, $^{171}Yb$ and $^{173}Yb$ [31], respectively), $\langle r^{-3} \rangle_{4f}$ is the expectation value of the $r^{-3}$ operator over a 4$f$-electron radial wave function equal to 11.07 and 12.5 at. units for $Er^{3+}$ and $Yb^{3+}$ [32], respectively, $Q$ is the quadrupole moment of the nucleus ($Q$=5.88 and 1.35 in units of $10^{-28}$ m$^2$ for $^{167}Er$ and $^{173}Yb$ [31], respectively), $\gamma_\infty$ = -80 and $R_Q$ = 0.1 are Sternheimer antishielding and shielding factors [33], $\hbar$ is the Planck constant. The first term in Eq. (4) comes from the crystal lattice contribution to the electric field gradient at the nucleus and contains the sum over the host lattice ions with charges $q_L$ and radius vectors $\mathbf{r}_L$ relative to the considered RE ion.

The CF energies of RE ions were found by numerical diagonalization of the operator $H_0 = H_{FI} + H_{CF}$. The matrices of the operators $H_Z$ and $H_{HF}$ were computed in the space of eigenfunctions of $H_0$, and the subsequent calculations were carried out with the Hamiltonian $H = E_0 + H_Z + H_{HF}$ (here $E_0$ is the diagonal matrix with elements equal to eigenvalues of $H_0$).

The average values of magnetic moments of RE ions versus the magnetic field **B** and temperature $T$ were computed in the local systems of coordinates accordingly to the definition

$$\langle \mathbf{\mu} \rangle_{\mathbf{B},T} = \text{Tr}\{\mu \exp[-(E_0 - \mathbf{\mu B})/k_BT]\}/\text{Tr}\{\exp[-(E_0 - \mathbf{\mu B})/k_BT]\}. \quad (5)$$

The obtained field dependences of the magnetization

$$M(\mathbf{B},T) = cN_A \sum_{n=1}^{4} \mathbf{B} < \mathbf{\mu}_n >_{\mathbf{B},T} /4B, \quad (6)$$

where the sum is taken over four nonequivalent RE sites in the unit cell, are compared with the experimental data in Figure 1. The magnetic susceptibility tensor in the local system of coordinates is diagonal and has only two different components $\chi_\perp = \chi_{XX} = \chi_{YY}$, $\chi_\parallel = \chi_{ZZ}$. The computed temperature dependences of the isotropic bulk magnetic susceptibilities $\chi(T) = [\chi_\parallel(T) + 2\chi_\perp(T)]/3$ of $Y_2Ti_2O_7$:$Er^{3+}$ and $Y_2Ti_2O_7$:$Yb^{3+}$ are compared with the results of measurements in Figure 2.

The EPR spectra corresponding to magnetic dipole transitions between sublevels of the well isolated ground doublet split by the external magnetic field can be described by the spin-Hamiltonian $H_S$ which represents the projection of the total Hamiltonian $H$ of an ion on the 2-dimensional ground state manifold. In the local system of coordinates, the spin-Hamiltonian is written as follows (we neglect nuclear Zeeman and quadrupole energies)

$$H_S = g_\parallel \mu_B B_z S_z + g_\perp \mu_B (B_x S_x + B_y S_y) + A_\parallel I_z S_z + A_\perp (I_x S_x + I_y S_y) \quad (7)$$

where the effective spin $S=1/2$, $g_{\parallel}$ and $g_{\perp}$ are components of the g-tensor, $A_{\parallel}$ and $A_{\perp}$ are the magnetic hyperfine constants. Exactly these spin-Hamiltonian parameters were determined from the analysis of the measured EPR spectra (see Table 1, columns "Exper."). The magnetic hyperfine interaction (3) can be written as follows: $H_{HFM} = \mathbf{aI}$, here the components of the vector $\mathbf{a}$

$$a_x = \mu_B \gamma_N \hbar \langle r^{-3} \rangle_{4f} \sum \left[ 2l_x - (O_2^0 - 3O_2^2)s_x + 3O_2^{-2}s_y + 6O_2^1 s_z \right], \quad (8)$$

$$a_y = \mu_B \gamma_N \hbar \langle r^{-3} \rangle_{4f} \sum \left[ 2l_y - (O_2^0 + 3O_2^2)s_y + 3O_2^{-2}s_x + 6O_2^{-1} s_z \right], \quad (9)$$

$$a_z = 2\mu_B \gamma_N \hbar \langle r^{-3} \rangle_{4f} \sum \left[ l_z + O_2^0 s_z + 3O_2^1 s_x + 3O_2^{-1} s_y \right] \quad (10)$$

operate in the space of electronic wave functions. The components of the g-tensor and the hyperfine constants in the spin-Hamiltonian are calculated using the expressions

$$g_{\parallel} = 2\langle +|kL_z + 2S_z|+\rangle, \quad g_{\perp} = 2\langle +|kL_x + 2S_x|-\rangle, \quad (11)$$

$$A_{\parallel} = 2<+|a_z|+>, \quad A_{\perp} = 2<+|a_x|-> \quad (12)$$

where $|+\rangle$ and $|-\rangle$ are the eigenfunctions of the Hamiltonian $H_0$ corresponding to the ground CF doublet of a Kramers ion. The equations (11,12) pave a bridge between the spin-Hamiltonian and crystal-field approaches. The calculated parameters of the spin-Hamiltonian (Table 1, columns "Theory") agree satisfactorily with the experimental data. We note that the obtained values of g-factors for $Yb^{3+}$ ions are close to those communicated in Ref. 15($g_{\parallel}$=1.79, $g_{\perp}$=4.27) for $^{170}Yb^{3+}$ ions in the concentrated system $Yb_2Ti_2O_7$. The estimated shifts of the hyperfine sublevels of the ground doublets of the $^{167}Er^{3+}$ and $^{173}Yb^{3+}$ ions induced by the quadrupole hyperfine interaction (4), which are less than 250 MHz, were not revealed in the measured spectra.

The relative integral intensities $W_{ij}$ of the resonant magnetic dipole transitions $i \to j$ between the eigen states of the spin-Hamiltonian (7) with energies $E_i$, $E_j$ and wave functions $|i\rangle$ and $|j\rangle$ induced by the microwave magnetic field $\mathbf{B}_1$ directed along the unit vector $\mathbf{e}$ perpendicular to the constant field $\mathbf{B}$ can be written as

$$W_{ij} = |<i|(g_{\perp}S_x e_x + g_{\perp}S_y e_y + g_{\parallel}S_z e_z)|j>|^2. \quad (13)$$

The spectral distribution of the absorption intensities for the fixed microwave frequency $\nu$ in the magnetic field $\mathbf{B}$ is approximated by the sum of Gaussians with the varied line width $\sigma$,

$$I(B) \sim \sum_{ij} W_{ij}(p_i - p_j) \frac{1}{\sqrt{2\pi}\sigma} \exp[-(\Delta_{ji} - h\nu)^2 / 2\sigma^2], \quad (14)$$

here $\Delta_{ji} = E_j(\mathbf{B}) - E_i(\mathbf{B})$, $p_i(\mathbf{B})$ is the population of the state $i$, and the sum over $i$ and $j$ is taken over all electron-nuclear sublevels of the electronic doublet in case of odd isotopes. As shown in Figures 3 and 4, the simulated spectral envelopes $dI(B)/dB$ reproduce well the registered EPR signals at different orientations of the magnetic fields.

## 4. Summary

We carried out detailed experimental and theoretical studies of static and dynamic magnetic properties of impurity $Er^{3+}$ and $Yb^{3+}$ ions in $Y_2Ti_2O_7$ single crystals. The crystal field approach was used to interpret the measured magnetic field and temperature dependences of the magnetization and the characteristics of the EPR spectra. The sets of the CF parameters determined earlier from the analysis of the optical spectra [24] allowed us to reproduce successfully the values of resonance fields and relative intensities of EPR signals for different orientations of the applied field and the magnetic anisotropy induced by strong fields in cubic pyrochlores $Y_2Ti_2O_7$:$Yb^{3+}$ and $Y_2Ti_2O_7$:$Er^{3+}$ containing small concentrations of 0.5 at.% of RE ions. The shapes of EPR lines including the fine hyperfine structures due to odd $^{171}Yb$, $^{173}Yb$ and $^{167}Er$ isotopes were successfully modeled and the corresponding line widths and parameters of hyperfine interactions were determined from the fitting procedure.

The obtained new information on single-ion spectral and magnetic properties of diluted RE pyrochlores can be used to revise parameters of the anisotropic exchange interactions in $Er_2Ti_2O_7$ and $Yb_2Ti_2O_7$ and to advance investigations of the intriguing magnetic properties of concentrated pyrochlores.

**Acknowledgments**

Financial support of the Russian Science Foundation under Grant № 19-12-00244is acknowledged.

**References**

1. Hallas A.M., Gaudet J., Gaulin B.D. *Ann. Rev. Condensed Matter Phys.* **9**, 105 (2018)
2. Rau J.G, Gingras M.J.P. *Annu. Rev. Condens. Matter Phys*. 10, 357 (2019)
3. Yan H., Benton O., Jaubert L., Shannon N. *Phys. Rev. B* **95**, 094422 (2017)
4. Champion J.D.M., Harris M.J., Holdsworth P.C.W., Wills A.C., Balakrishnan G., Bramwell S.T., Čižmár E., Fennell T., Gardner J.S., Lago J., McMorrow D.F., Orendáč M., Orendáčová A., Paul D.McK., Smith R.I., Telling M.T.F., Wildes A. *Phys. Rev. B* **68**, 020401 (R) (2003)
5. Poole A., Wills A.S., Lelievre-Berna E. *J. Phys.: Condens. Matter* **19**, 452201 (2007)
6. Dun Z.L., Li X., Freitas R.S., Arrighi E., Dela Cruz C.R., Lee M., Choi E.S., Cao H.B., Silverstein H.J., Wiebe C.R., Cheng J.G., Zhou H.D. *Phys. Rev. B* **92**, 140407 (2015)
7. Hallas A.M., Gaudet J., Wilson M.N., Munsie T.J., Aczel A.A., Stone M.B., Freitas R.S., Arevalo-Lopez A.M., Attfield J.P., Tachibana M., Wiebe C.R., Luke G.M., Gaulin B.D. *Phys. Rev. B* **93**, 104405 (2016)
8. Petit S., Lhotel E., Damay F., Boutrouille P., Forget A., Colson D. *Phys. Rev. Lett.* **119**, 187202 (2017)
9. Hallas A.M., Gaudet J., Butch N.P., Xu G., Tachibana M., Wiebe C.R., Luke G.M., Gaulin B.D. *Phys. Rev. Lett.* **119**, 187201 (2017)
10. Yasui Y., Soda M., Iikubo S., Ito M., Sato M., Hamaguchi N., Matsushita T., Wada N., Takeuchi T., Aso N., Kakurai K. *J. Phys. Soc. Jpn.* **72**, 3014 (2003)
11. Chang L.-J., Onoda S., Su Y., Kao Y.-J., Tsuei K.-D.,Yasui Y., Kakurai K., Lees M.R. *Nature Commun.* **3**, 992 (2012)
12. Yaouanc A., Dalmas de Reotier P., Bonville P., Hodges J. A., Glazkov V., Keller L., Sikolenko V.,Bartkowiak M., Amato A., Baines C., King P.J.C., Gibbens P.C.M., Forget A. *Phys. Rev. Lett.* **110**, 127207 (2013)
13. Yasui Y., Hamachi N., Kono Y., Kittaka S., Sakakibara T. *SPIN* **5**, 1540002 (2015)
14. Yaouanc A., de Reotier P. D., Keller L., Roessli B., Forget A.*J. Phys.: Condens. Matter* **28**, 426002 (2016)
15. Hodges J.A., Bonville P., Forget A., Rams M., Królas K., Dhalenne G. *J. Phys.: Condens. Matter* **13**, 419301 (2001)
16. Cao H., Gukasov A., Mirebeau I., Bonville P., Decorse C., Dhalenne G. *Phys. Rev. Lett.* **103**, 056402 (2009)
17. Malkin B.Z., Zakirov A.R., Popova M.N., Klimin S.A., Chukalina E.P., Antic-Fidancev E., Goldner Ph., Aschehoug P., Dhalenne G. *Phys. Rev. B* **70**, 075112 (2004)
18. Gaudet J., Maharaj D.D., Sala G., Kermarrec E., Ross K.A., Dabkowska H.A., Kolesnikov A.I., Granroth G.E., Gaulin B.D. *Phys. Rev. B* **92**, 134420 (2015)
19. Sala G., Maharaj D. D., Stone M. B., Dabkowska H. A., Gaulin B. D. *Phys. Rev. B* **97**, 224409 (2018)
20. Gaudet J., Hallas A.M., Kolesnikov A.I., Gaulin B.D. *Phys. Rev. B* **97**, 024415 (2018)
21. Gardner J.S., Ehlers G. *J. Phys.: Conens. Matter* **21**, 436004 (2009)
22. Malkin B.Z., Lummen T.T.A., van Loosdrecht P.H.M., Dhalenne G., Zakirov A.R.*J. Phys.: Condensed Matter* **22**, 276003 (2010)

23. Bertin A., Chapuis Y., Dalmas de Réotier P., Yaouanc A. *J. Phys.: Condens. Matter* **24**, 256003 (2012)
24. Batulin R.G., Cherosov M.A., Gilmutdinov I.F., Khaliulin B.F., Kiiamov A.G., Klekovkina V.V., Malkin B.Z., Mukhamedshin I.R., Mumdzhi I.E., Nikitin S.I., Rodionov A.A., Yusupov R.V. *Physics of the Solid State* **61**, 818 (2019)
25. Li Q.J., Xu L.M., Fan C., Zhan F.B., Lv Y.Y., Ni B., Zhao Z.Y., Sun X.F. *J. Cryst. Growth* **377**, 96 (2013)
26. Bain G.A., Berry J.E. *J. Chem. Education* 85, 532 (2008)
27. Cao H.B., Gukasov A., Mirebeau I., Bonville P., Decorse C., Dhalenne G. *Phys. Rev. Lett.* **103**, 056402 (2009)
28. Sosin S.S., Prozorova L.A., Lees M.R., Balakrishnan G., Petrenko O.A. Phys. Rev. B 82, 094428 (2010)
29. Crosswhite H.M., Crosswhite H. *J. Opt. Soc. Am. B* **1**, 246 (1984)
30. Klekovkina V.V., Zakirov A.R., Malkin B.Z, Kasatkina L.A. *J. Phys.: Conf. Ser.* **324**, 012036 (2011)
31. McCausland M.A.H., Mackenzie I.S. *Adv. Phys.* **28**, 305 (1979)
32. Abragam A., Bleaney B. *Electron Paramagnetic Resonance of Transition Ions*, Clarendon Press, Oxford, UK (1970)
33. Gupta R.P., Sen S.K. *Phys. Rev. A* **7**, 850 (1973)

**Table 1.** The spin-Hamiltonian parameters for impurity $Yb^{3+}$ and $Er^{3+}$ ions in $Y_2Ti_2O_7$. Calculated without hybridization with the charge transfer states values of g-factors $Yb^{3+}$ ions are marked by stars.

|  | $^{171}$Yb | | $^{173}$Yb | | $^{167}$Er | |
|---|---|---|---|---|---|---|
|  | Exper. | Theory | Exper. | Theory | Exper. | Theory |
| $g_\parallel$ | 1.787 | 1.864 | 1.787 | 1.864 | 2.29 | 2.305 |
| $g_\perp$ | 4.216 | 4.181 | 4.216 | 4.181 | 6.76 | 6.811 |
| $A_\parallel$ (MHz) | 1235 | 1353 | −353.8 | −388 | −277.4 | −252 |
| $A_\perp$ (MHz) | 3064 | 3050 | −881.4 | −875 | −810.9 | −735 |

**Table 2.** CF parameters $B_p^k$ (cm$^{-1}$) for the impurity $RE^{3+}$ ions in $Y_2Ti_2O_7$:$Er^{3+}$ and $Y_2Ti_2O_7$:$Yb^{3+}$.

| p k | 2 0 | 4 0 | 4 3 | 6 0 | 6 3 | 6 6 |
|---|---|---|---|---|---|---|
| $Er^{3+}$ | 239.8 | 311.8 | −2305.2 | 45.7 | 666.6 | 753.2 |
| $Yb^{3+}$ | 264.8 | 270.8 | −2155.2 | 44.9 | 636.6 | 683.2 |

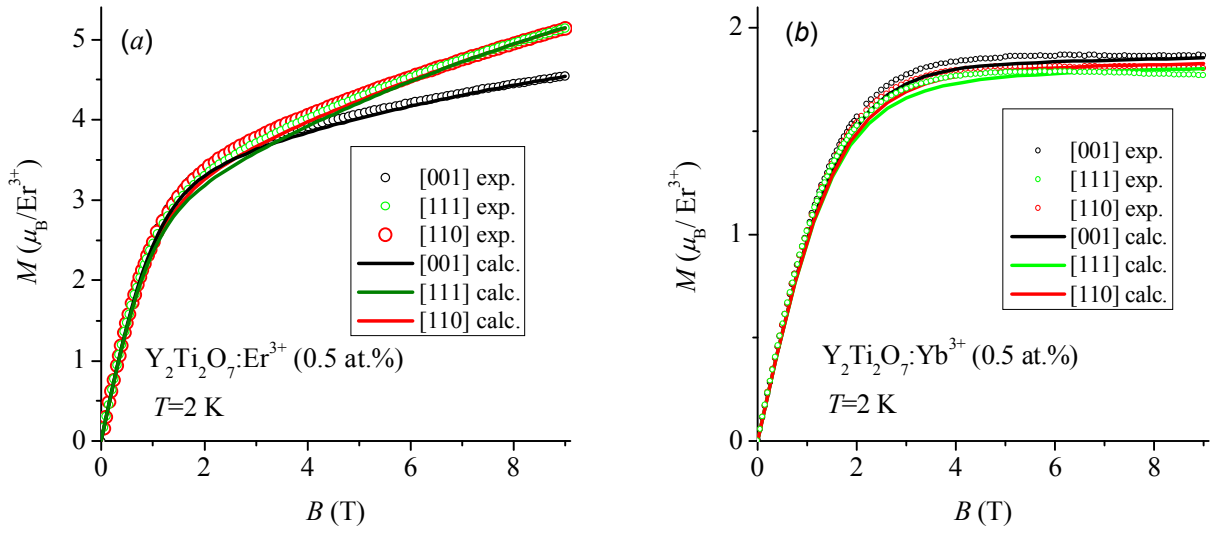

**Figure 1.** Measured (symbols) and calculated (solid curves) field dependences of the magnetization of $Y_2Ti_2O_7$:$Er^{3+}$ (*a*) and $Y_2Ti_2O_7$:$Yb^{3+}$ (*b*) single crystal for magnetic fields *B* directed along the crystallographic axes at temperature *T*=2 K.

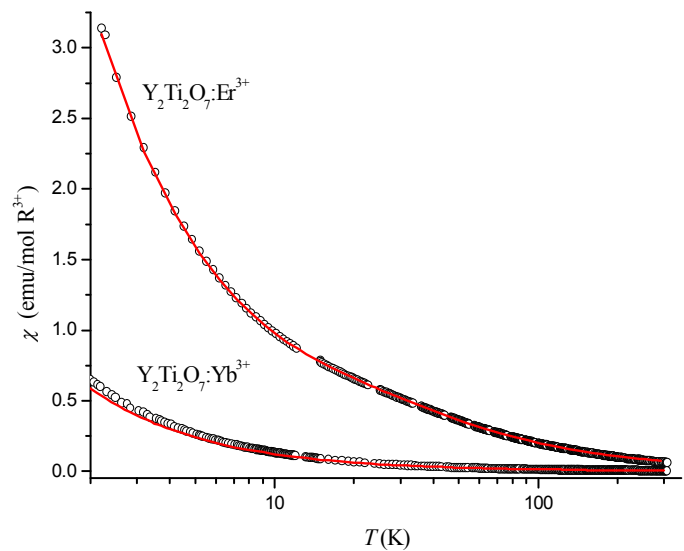

**Figure 2**. Measured (symbols) and calculated (solid curves) temperature dependences of the dc-susceptibilities of of $Y_2Ti_2O_7:R^{3+}$ (R=Er, Yb, 0.5 at.%) single crystals.

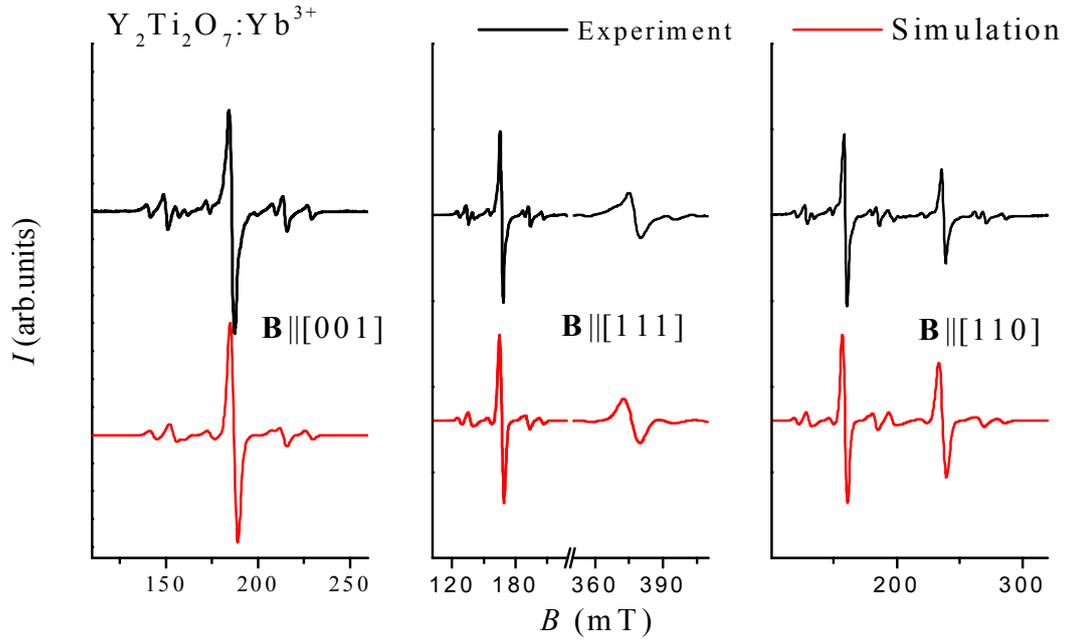

**Figure 3.** EPR spectra of the $Y_2Ti_2O_7$:$Yb^{3+}$ (0.5 at.%) single-crystal for three orientations of the magnetic field $B_0$ ($\upsilon$ =9.41553 GHz, $T$=15 K). Experimental data and the results of simulations are shown by solid black and red lines, respectively. Linewidths of the spectral components of the simulated spectra had the values of $\sigma$=0.105 GHz for **B**∥[001]), 0.12 GHz for **B**∥[110]), and $\sigma_1$=0.09 GHz and $\sigma_2$=$\sigma_3$=$\sigma_4$=0.12 GHz for **B**∥[111]

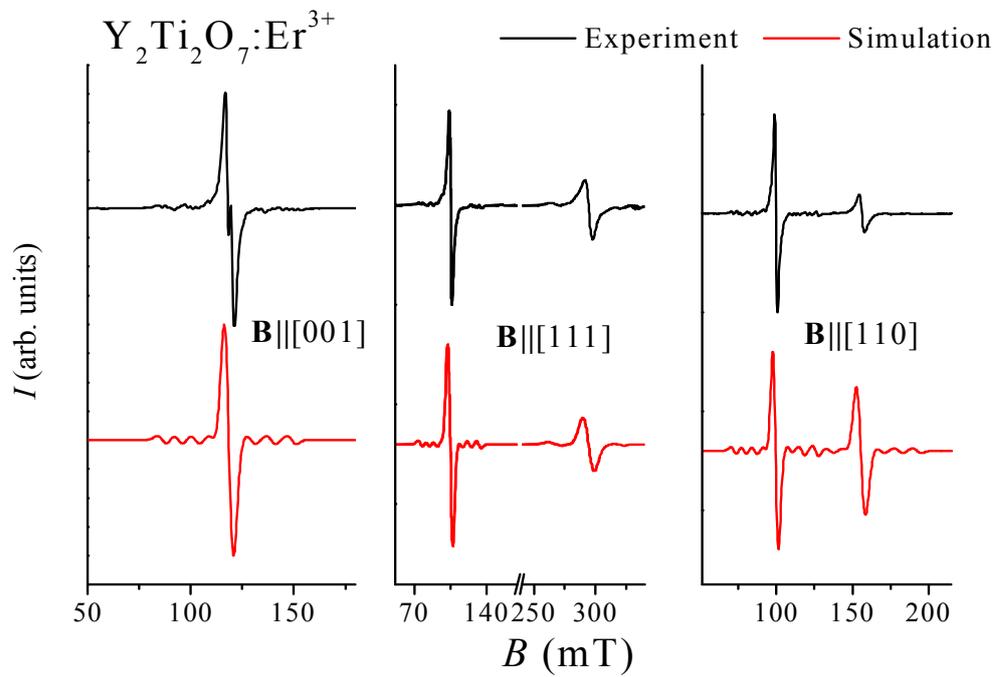

**Figure 4.** EPR spectra of $Y_2Ti_2O_7$:$Er^{3+}$ (0.5 at.%) single-crystal for three orientations of the magnetic field $B_0$ ($\upsilon$ =9.41553 GHz, $T$ = 20 K). Experimental data and the results of calculations are shown by solid black and red lines, respectively. Linewidths in the simulations had the values of σ=0.18 GHz for **B** ||[001] and **B** ||[110], $\sigma_1$=0.15 GHz and $\sigma_2$=$\sigma_3$=$\sigma_4$=0.225 GHz for **B** ||[111].